\documentclass{article}
\usepackage[utf8]{inputenc}
\usepackage{geometry}
\usepackage{graphicx}
\usepackage{xcolor}
\usepackage{bm}
\usepackage{amsmath}
\usepackage[normalem]{ulem}
\usepackage{graphicx}
\usepackage{color,soul}
\usepackage[colorlinks=true, citecolor=black, urlcolor=blue]{hyperref}
\usepackage{authblk}

\begin{document}

\title{Machine Learning for Multi-fidelity Scale Bridging and Dynamical Simulations of Materials}

\author[1]{Rohit Batra}
\author[1,2]{Subramanian Sankaranarayanan}
\affil[1]{Center for Nanoscale Materials, Argonne National Laboratory, Lemont, Illinois 60439, United States}
\affil[2]{Department of Mechanical and Industrial Engineering, University of Illinois,Chicago, Illinois 60607, United States}

\date{\today}
\maketitle
\begin{abstract}
    Molecular dynamics (MD) is a powerful and popular tool for understanding the dynamical evolution of materials at the nano and mesoscopic scales. There are various flavors of MD ranging from the high fidelity albeit computationally expensive \textit{ab-initio} MD to relatively lower fidelity but much more efficient classical MD such as atomistic and coarse-grained models. Each of these different flavors of MD have been independently used by materials scientists to bring about breakthroughs in materials discovery and design. A significant gulf exists between the various MD flavors\textcolor{black}{, each having varying levels of fidelity. The accuracy of DFT or \textit{ab-initio} MD is generally much higher than that of classical atomistic simulations which is higher than that of coarse-grained models.} Multi-fidelity scale bridging to combine the accuracy and flexibility of \textit{ab-initio} MD with efficiency classical MD has been a longstanding goal. The advent of big-data analytics has brought to the forefront powerful machine learning methods that can be deployed to achieve this goal. Here, we provide our perspective on the challenges in multi-fidelity scale bridging and trace the developments leading up to the use of machine learning algorithms and data-science towards addressing this grand challenge.
\end{abstract}

\section{Introduction}
Ever since the first successful demonstration of molecular dynamics (MD) to determine the diffusion coefficient of liquid argon more than 50 years ago \cite{rahman1964correlations}, MD has become an indispensable tool to understand and predict materials properties for a broad range of applications, including drug discovery\cite{de2016role, liu2018molecular, frederix2015exploring, hollingsworth2018molecular}, materials design\cite{sresht2015liquid, singh2015computational, belmares2004hildebrand} and defect chemistry\cite{bagri2011thermal, skoulidas2005self, luo2012enhancement}. Its generality, i.e., being applicable to a diverse range of materials (metals, ceramics, amorphous glasses, polymers, or biomolecules), along with its versatility, i.e, the ability to capture thermodynamic, mechanical, electrical and chemical behavior of materials, make it a pervasive and vital theoretical tool, as highlighted in Figure 1a. Increasingly, empirical findings are supported with \textcolor{black}{analogous} MD simulations to explain materials behavior at the electronic, atomistic or molecular level \cite{hura2003water,chen2018thermal}.
Beyond its scientific merit, the popularity of MD simulations is fueled by the advancements in the computational power, and the availability of efficient software packages (VASP\cite{hafner2008ab}, LAMMPS \cite{plimpton1993fast}, ASE\cite{larsen2017atomic}, NAMD \cite{phillips2005scalable}, CHARMM \cite{brooks2009charmm}, GROMACS \cite{berendsen1995gromacs}, etc.) that allow even a novice user \textcolor{black}{(with limited background in MD simulations)} to quickly simulate their experiments virtually. However, depending on the the phenomenon or material (metals, polymers, biomolecules or proteins) being modeled, smaller MD sub-communities have emerged to address specific modeling needs.
Machine learning (ML) based force fields (FF), which is the focus of this perspective, provide an opportunity to bridge distinction between these different sub-communities, and provide a common and flexible methodology applicable across diverse materials domain.

At its core, classical MD simulations uses a physical model typically termed `force-fields' to iteratively compute inter-atomic forces, \textcolor{black}{which when used to integrate Newton's equations of motion using numerical methods such as velocity Verlet for all atoms of a system}, can reveal its dynamics. Depending on the flavor of the model used to obtain these inter-atomic forces, i.e., {\it ab-initio}, classical or coarse-grained, the nature of the simulations can vary drastically. First-principles methods, such as density functional theory (DFT) \cite{hohenberg1964inhomogeneous, kohn1965self}, provide the most general and reliable force estimate but are computationally too exhaustive to be able to simulate
phenomenon that involves larger system sizes (tens of nanometers to microns and beyond for understanding mechanical failure, microstructure evolution, to name a few) and/or occurs over longer time scales (nanoseconds and beyond for phenomena such as nucleation and growth, fatigue, to name a few).
The empirical FFs or classical \textcolor{black}{inter-atomic} potentials community have successfully surmounted this challenge by developing exclusive analytical functional forms, involving \textcolor{black}{tens to hundreds} of tunable parameters, to approximate the inter-atomic force (or energy) expression. Once these potentials are fit to a known set of thermodynamic, mechanical or physical properties, they can be utilized to simulate materials dynamics under different conditions of interest.   To efficiently treat even larger system sizes, such as biomolecules or proteins, coarse-grained representations are necessary, wherein rather than an all atomistic treatment, molecular motifs form the basic building block. \textcolor{black}{In classical atomistic molecular dynamics, it should be noted that the terms force-fields and inter-atomic potentials are often interchangeably used to describe the functional form that captures the interactions between the atoms in the system. For coarse-grained simulations, the term force-field is used since no explicit atoms are considered.} Figure 1b illustrates the different time/length scale regimes treated using aforementioned theoretical models.

\begin{figure}
	\centering 
	\includegraphics[width=1\textwidth]{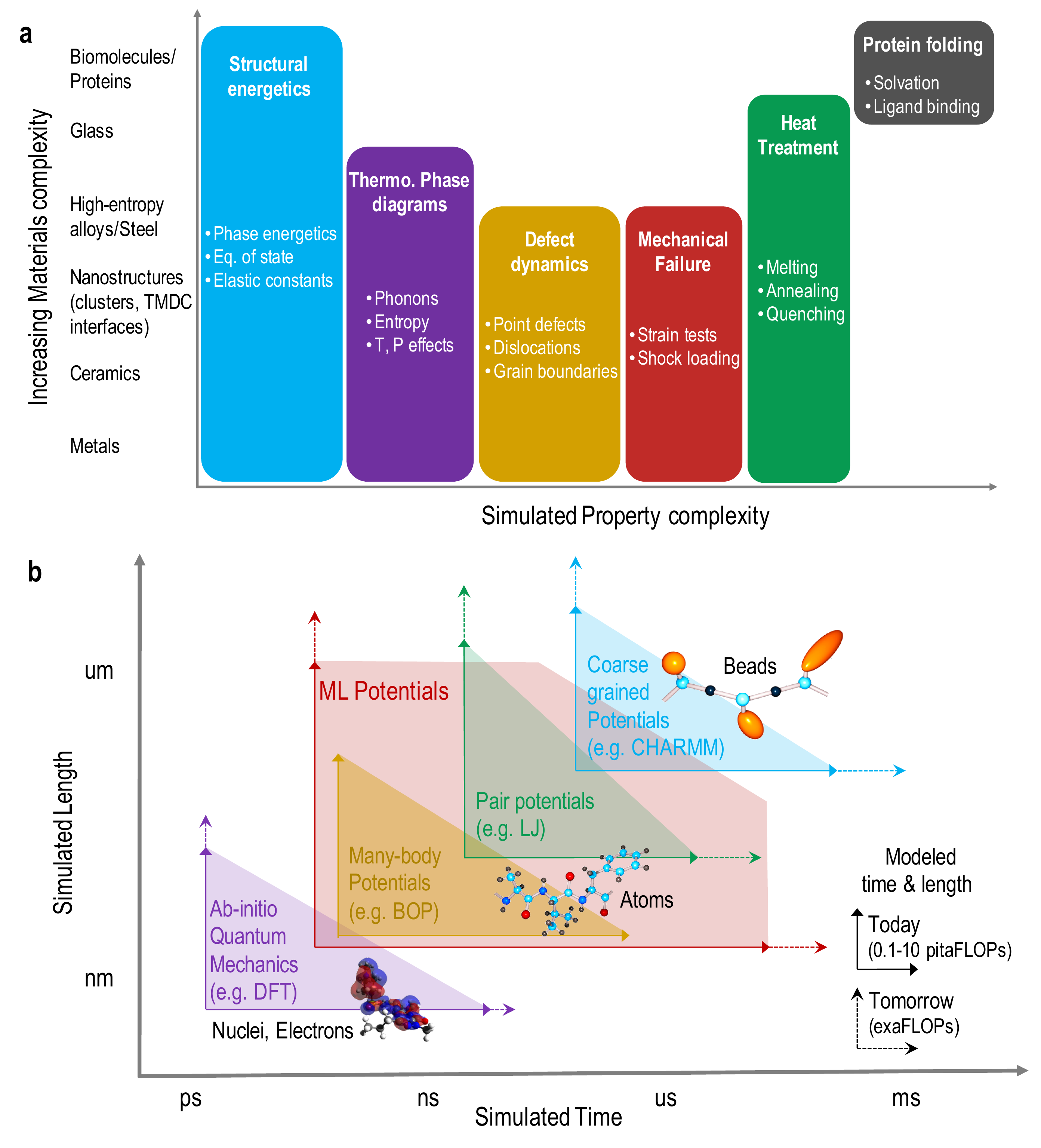}
	\caption{(a) The diverse materials and property space that can be treated using MD simulations. (b) Different theoretical models employed to treat atomic interactions during MD simulations, along with their fundamental building blocks and region of applicability. Depending on the choice of atomic fingerprint ML potentials can vary in computational cost significantly. \textcolor{black}{Note that 'tomorrow' refers to the increase in the (future) computational resources that can enable  an increase in time and length scale that can be captured by MD simulations.} }
	\label{fig:figure1}
\end{figure}
    
Owing to the simplicity of the force expression, empirical FFs or coarse-grained models are several orders of magnitude faster than {\it ab-initio} molecular dynamics (AIMD)
\cite{car1985unified}. However, both of these former approaches suffer from limited accuracy and transferability when applied for cases far from those considered during the parameter fitting process. \textcolor{black}{There thus exists a significant gulf between the various flavors of molecular dynamics ranging from the high-fidelity yet high computational cost \textit{ab-initio} MD to low-fidelity yet efficient coarse-grained MD.} Perhaps, an even bigger deterrent for FFs with fixed functional forms is the years of human effort required to find a respectable set of parameters, referred to as FF parameterization, that reliably describes the physics of a specific material system.


The advent of big data analytics and machine learning has brought to the forefront a new generation of powerful atomistic and molecular models \cite{chan2019machinec, kim2018polymer, mueller2016machine, butler2018machine, ward2017atomistic, chandrasekaran2019solving, pilania2017multi}. This has recently propelled the development of AI-assisted or machine learning based FF that have displayed the promise of bridging the accuracy(fidelity)/flexibility gap between the AIMD and classical/coarse-grained models, while being sufficiently rapid to reach larger length and time scales. \textcolor{black}{Machine learning and AI are playing a critical role towards addressing this challenge in multi-fidelity scale bridging.} Furthermore, these FFs can be built relatively quickly and in an autonomous manner, without the need of human intervention.

In this perspective, we discuss two general directions in which the ML methods are emerging as indispensable tools for the MD community: 1) efficient and autonomous parameterization of FFs with known or pre-defined functional forms, and 2) directly learning the FF functional from the available high-fidelity AIMD or \textcolor{black}{experimental} datasets (i.e., establishing a configuration-to-energy/force mapping). We discuss several examples from each areas, along with their advantages and limitations, and finally conclude with some research avenues that remain to be addressed in the future.

\section{Machine Learning for Parameterization of Pre-defined Force-Fields}
Historically, FFs or potentials have been constructed by fitting an approximate atomic energy (or force) expression to reproduce tabulated empirical (lattice parameters, elastic constants, melting point, etc.) or quantum mechanical (phase energetics, cluster and interface energies, atomic forces, etc.) data. The complexity or functional form of the atomic energy expression is chosen based on the types of atomic interactions (ionic, covalent, dispersive, hydrogen bonding, and electrostatic) believed to be dominant in the considered material system, with the number of fitting parameters varying from two---Lennard-Jones potential \cite{jones1924determination} for simpler noble elements---to nearly 100---ReaxFF\cite{chenoweth2008reaxff} and COMB \cite{shan2010charge} potentials for complex multi-elemental or reactive systems. The ingenuity of a FF developer lies in not only choosing an appropriate functional form for the material system of interest, but perhaps more importantly, in tuning the associated parameters to reproduce the known material behavior. Unfortunately, little guidance is available to solve this high-dimensional FF parameterization problem, with developers mostly relying on ``chemical intuition" \textcolor{black}{(in simple words - the expertise of the computational chemist)} or traditional heuristics. Moreover, for complex systems (glasses, reactions, etc.) the number of tunable parameters increases dramatically and such heuristics become difficult to implement. Thus, there is no surprise that FF parameterization is considered to be a very painful exercise, often entailing several years of research efforts, and is pursued by only a limited number of scientists.




ML and AI based tools, however, offer an efficient and autonomous alternative pathway for FF parameterization, especially for complex systems involving large number of tuning parameters. Two methods that have particularly shown promise in this area are the Bayesian optimization (BO) \cite{mockus2012bayesian} and the genetic algorithm (GA) \cite{mitchell1998introduction}. BO can be used for FF parameterization by coupling it with a supervised ML model (e.g. Gaussian process regression, GPR \cite{rasmussen2003gaussian}) that predicts the performance of a given set of FF parameters with some uncertainty. The main idea behind the use of BO is to find an optimal set of parameters that offers the best ``exploration versus exploitation trade-off," that is, the best balance between exploring \emph{new} parameter space where the ML model is uncertain about the FF performance, and on the other hand focusing the search in the region where the ML model predicts FF parameters to have low errors. By iteratively following this search strategy, FF parameters can be found that minimizes the parameterization cost function, as illustrated in Figure 2a. Bauchy and co-workers used this approach to develop FFs for glassy silica \cite{liu2019parameterization}, a material system known to be difficult to parameterize using traditional methods. They combined BO with GPR to tune interatomic potential of the Buckingham form that involves ten independent parameters and pose a difficult high-dimensional optimization problem.
FF performance was evaluated using a cost function measuring the difference in the partial pair density functions (PDF) of Si and O obtained from the FF and the baseline AIMD simulations.
To initiate the BO process, they first trained a GPR model that predicts the cost function, given the FF parameters. Then using the expected improvement \cite{movckus1975bayesian} scheme they iteratively selected the new set of FF parameters that could potentially lead to reduction in the cost function. After only 600 iterations they were able to find a good FF, which not only reproduced the correct Si and O PDFs, but was significantly different from those of the baseline BKS potential—which illustrates the roughness of the cost function, and the challenge with traditional FF development approach. Notably, this scheme is general and can be extended to develop FFs for other material systems.

Another approach that has shown promise to parameterize FFs are evolutionary optimization methods such as GA, which solves the optimization problems using the principle of natural selection. In analogy with how nature uses the basic steps of \emph{crossover}, \emph{mutation} and \emph{selection} for evolution of a specie, GA can be used to evolve a generation of FF parameters to reduce the errors in the parameterization cost function. In this approach, starting from a randomly selected generation of (nearly 100) FF parameter candidates, 1) crossover---mixing parameter values among different sets---and mutation---randomly perturbing parameter values---operations are performed to obtain new candidate sets; 2) MD computations are then performed to evaluate performance of these candidates using the parameterization cost function; 3) top candidates with low cost evaluations are then retained as parents for the next generation of parameter sets, with the above steps iterated until a candidate parameter set with desired performance is obtained. Although the GA approach offers no theoretical convergence guarantees \cite{takahashi1998convergence, corne2018evolutionary}, empirically it has been shown to perform extremely well, especially for problems involving rugged cost function with several deceptive low-value local minimas. For example, GA has been used to fit ReaxFF for SiOH system \cite{larsson2013global}.

A major bottleneck in the use of GA (or another optimization scheme) for FF parameterization is the evaluation of the cost function (step 2) for each candidate parameter set, which involves performing several expensive MD computations (to estimate correct phase energetics, high-temperature high-pressure stability, etc.). To overcome this problem, Chan and co-workers used a hierarchical objective function scheme, termed HOGA, presented in Figure 2b \cite{chan2019machinea, chan2019machineb, loeffler2019teaching}. In this approach, the desired set of property objectives are segregated into different hierarchical classes based on their computational cost. During the cost function evaluation (i.e. step 2), the processes is truncated if a candidate parameter set leads to large errors in a hierarchical classes, with the associated class penalty being assigned to the candidate set. Thus, the HOGA scheme accelerates the traditional evolutionary search by (i) efficiently sampling the parameter landscape in a given generation, and (ii) overcomes the limitation of a single objective that relies on assigning arbitrary weights. Further, it allows multi-fidelity scale bridging by incorporating training datasets at various length scales, i.e., phase energetics and structural properties from first principles or atomistic models, experimental measurements, and even on-the-fly MD based property evaluations. The sequencing and selection of hierarchical classes is at the discretion of the user, which can be modified based on the target property of interest. The success of the HOGA scheme was demonstrated by developing FFs for H$_2$O \cite{chan2019machinea} and WSe$_2$ \cite{chan2019machineb} systems using bond order potential model with 13 independent parameters. The coarse-grained (CG) models for water accurately captures its density anomaly, which has been a major limitation for other CG models. Similarly, WSe$_2$ FFs were used to study thermal transport behavior of diverse nanostructures at different temperatures using non-equilibrium MD simulations. In contrast to the traditional local optimization approaches, these evolutionary optimization methods allow for global optimization of the objective function. The success of these evolutionary approaches stems from their ability to more efficiently navigate the high-dimensional parameter space than conventional local optimizers or gradient based approaches \cite{corne2018evolutionary}.

\begin{figure}
	\centering 
	\includegraphics[width=1\textwidth]{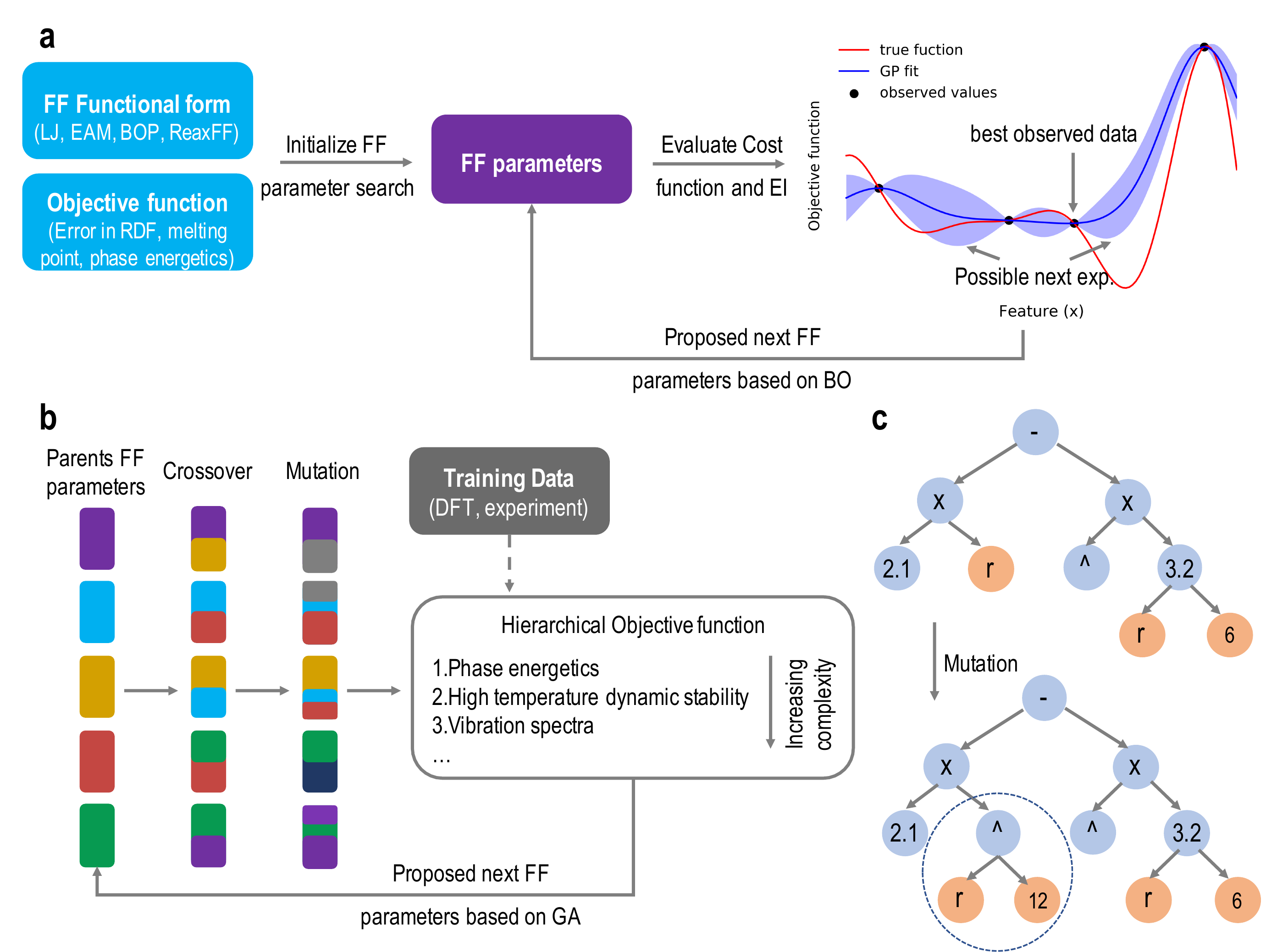}
	\caption{ML schemes for efficient (and autonomous) parameterization of FFs (a) Bayesian optimization, and (b) hierarchical objective function scheme based on GA (HOGA). (c) Tree structure representation of inter-atomic potentials. An example mutation operation resulting in recovery of the Lennard-Jones potential using symbolic regression.}
	\label{fig:figure2}
\end{figure}

Although BO and GA provide efficient ways to parameterize FFs, a key aspect that remains missing from the whole process of FF development is the generation of training data. In this regards, Chan and co-workers developed an autonomous framework termed BLAST---Bridging Length/Time scales via Atomistic Simulation Toolkit---that allows users to create their own potentials/FFs by following through each stage of FF development, i.e., generating appropriate training data sets, optimizing FF parameters, and finally cross-validating their FF predictions {\cite{chan2019machinec}}. It employs a data-driven and ML based FF development approach that significantly deviates from the traditional approaches in several key aspects, including the nature of the training dataset, the use of ML algorithms for advanced sampling and FF parameter optimization (e.g., genetic algorithms, multi-objective optimization, neural networks, etc.), the available form of FFs, and cross validation and iterative improvement of ML based potentials. This approach has been successfully demonstrated for diverse material systems, including metal nanoclusters, oxides, nitrides, and 2D materials (stanene and WSe2){\cite{narayanan2016multi,narayanan2016gold,fatih2014iridium,chan2019machinec,chan2019machineb,cherukara2016stanene}}.

\textcolor{black}{Finally, we note the following in the area of FF parameter optimization using machine learning. Although optimization procedures such as BO and GA remain attractive to optimize the coefficients or parameters of any given FF, we envision the use of other emerging AI techniques, for instance, Monte Carlo Tree Search (MCTS) \cite{silver2016mastering} with reinforcement learning to derive globally optimal solutions for these parameter search problems. We expect burgeon of automated workflows (e.g. BLAST - Bridging Length/timescales via Atomistic Simulation Toolkit \cite{chan2019machinec}) that allow non-expert users to carry out their own potential model development depending on the target system and phenomenon. The term 'workflow' describes the use of scripting languages to interface with various domain-specific simulation codes and/or execution of simulation campaigns across multiple HPC and cloud resources. These can be achieved via python or dedicated parallel scripting languages, such as SWIFT \cite{wilde2011swift}. Likewise, other dataflow systems like Decaf developed by Tom Peterka allow for the parallel communication of coupled tasks in an HPC workflow \cite{dreher2017decaf}. This new development democratizes the FF development process and reduces the dependency of the MD community on a handful of FF developers for optimization of FF parameters. Furthermore, this approach of FF development is believed to remain popular as it returns a FF functional form which is physically meaningful, interpretable and much faster than a complete ML based FF, which is discussed next.}

\section{Force Fields with Machine Learned Functional Forms}

Despite the popularity of pre-defined functional forms, it has become imperative to address the limitations arising from their lack of flexibility. The use of pre-defined functions to represent the inter-atomic interactions inherently imposes a limit on its predictive power. This has led to the search for alternate more flexible, accurate and efficient models. Rather than using ML to find parameters of a FF with fixed functional form (bond order potential (BOP) \cite{finnis2007bond}, reactive force-fields ReaxFF \cite{chenoweth2008reaxff}, etc.), therefore, a different class of ML FFs have emerged in the last decade that aim at directly learning the functional form of the FF itself, i.e. establish a direct mapping from the atomic neighborhood, or \emph{fingerprint}, to atomic energies---and their sum to total potential energy---using the reference quantum mechanical data. As no prior restriction is imposed on the functional form of these FFs, this methodology is quite general and can be used to learn atomic energy functionals of diverse materials (metals, ceramics, alloys, polymers) involving different atomic interactions with minimal human interference. This also precludes the limitations of the traditional FFs with fixed functional form designed for capturing specific inter-atomic interactions.
This approach has been successfully used to develop ML FFs for numerous systems, including elemental bulk (e.g. Al \cite{botu2017machine, huan2017universal, kuritz2018size}, C \cite{rowe2018development, podryabinkin2019accelerating, khaliullin2010graphite}, Li \cite{podryabinkin2017active}, Si\cite{behler2007generalized, behler2008metadynamics, huan2017universal, deringer2018realistic}, Fe \cite{dragoni2018achieving}, Zr \cite{zong2018developing}), alloys\cite{artrith2015grand}, metallic clusters \cite{chiriki2016modeling, chiriki2017neural}, semiconductors \cite{sosso2012neural}, oxides \cite{artrith2011high, artrith2016implementation}, water \cite{morawietz2016van, cheng2016nuclear}, organic molecules \cite{jose2012construction, gastegger2016comparing}, among others, and study complex applications, such as surface diffusion \cite{boes2017neural}, liquids, phase equilibria, etc. Different software packages (RuNNer \cite{behler2007generalized}, \textcolor{black}{GAP suite} \cite{bartok2010gaussian}, AMP \cite{khorshidi2016amp}, AGNI \cite{botu2017machine}, DeepMD \cite{wang2018deepmd}, AENet \cite{artrith2016implementation}, MTP \cite{shapeev2016moment}, SchNetPack \cite{schutt2018schnetpack}, N2P2 \cite{desai2020implementing}, \textcolor{black}{SNAP} \cite{thompson2015spectral}, etc.) that allow users to train and validate their own ML FF for a particular system of interest have also been released. Despite its popularity and success over the past several years, it is still an active area of research with advancements in atomic fingerprinting scheme or the supervised ML task (atomic positions to energy functional mapping) being reported frequently. Thus, below we provide only an essence of this approach, and refer the reader to detailed reviews on this topic for more information\cite{bartok2013representing, rupp2015machine, behler2011neural, deringer2019machine, behler2017first, ramprasad2017machine, behler2016perspective, handley2010potential, mueller2016machine, chan2019machinec}.

Figure 3 shows a general schema typically used in the construction of these ML based FFs that consists of three stages: database generation, fingerprinting and supervised learning. In the first stage a reference high-fidelity first-principles (e.g. AIMD) database spanning diverse material configurational space and associated energetics, stresses and atomic forces is constructed. In the next step, a \emph{fingerprinting} scheme is utilized to numerically represent the atomic configurations present within the reference dataset, which are then mapped to the associated atomic energies---or more formally, their sum to the total potential energy---in the last step using a supervised ML algorithm. Within each of these three stages, numerous methods have been proposed to improve their accuracy or speed, thereby creating different flavors of ML FFs in the community.

Perhaps the most critical component in the construction of these ML based FFs has been the development of a suitable atomic fingerprint in stage 2, as these fingerprints are required to be strictly invariant with respect to arbitrary translations, rotations, and exchange of like atoms, in addition to being continuous and differentiable with respect to small variations in atomic positions. Several candidates, including those based on atom-centered symmetry functions (ACSF) \cite{behler2007generalized}, smooth overlap of atomic positions (SOAP)\cite{bartok2010gaussian, szlachta2014accuracy, bartok2015gaussian}, moment tensor representation \cite{shapeev2016moment}, spectrum of London and Axilrod–Teller–Muto (SLATM) \cite{huang2016communication}, the Faber–Christensen–Huang–Lilienfeld (FCHL) representations \cite{faber2018alchemical, christensen2020fchl}, bispectra of neighborhood atomic densities \cite{thompson2015spectral}, Coulomb matrices (and its variants) \cite{rupp2012fast, chmiela2017machine}, and others, have been proposed. A common theme across all of these fingerprinting approaches is to numerically capture atomic neighborhood around a reference atom using some form of distribution functions as qualitatively illustrated in Figure 3a. The accuracy vs speed trade-off among these different descriptors is being presently studied \cite{zuo2019performance}, but the ACSF and SOAP representations have been most successful in terms of number of applications. We, however, note that other descriptors have also been shown to produce reliable FFs, hinting that many of these representations maybe accurate enough for ML FF generation.

In terms of the first data generation step, researchers have proposed different MD or sampling techniques to generate high-quality diverse database with limited computational budget \cite{deringer2018data, deringer2017machine, hajinazar2017stratified, huan2017universal}. Learning algorithms, ranging from neural networks \cite{behler2007generalized}, support vector machines \cite{balabin2011support}, kernel ridge regression \cite{huan2017universal, rupp2012fast}, GPR \cite{bartok2010gaussian}, etc. have been utilized to carry out the supervised learning task in stage 3. Although it is not clear which supervised learning algorithm performs best, deep neural networks are expected to be a top choice because of the data intensive nature of this problem.

A relatively new development within the area of ML FFs is to learn and predict the atomic forces directly \cite{li2015molecular, botu2017machine, glielmo2017accurate}, as this is the fundamental quantity required to perform MD simulations. These approaches are inspired by the idea that it should be possible to predict atomic forces given just the atomic configuration, without going through the agency of the total potential energy. A major advantage of this approach is that atomic force can be uniquely assigned to an individual atom, while the potential energy is a global property of the entire system; partitioning the potential energy to atomic contributions does not have a formal basis, but is often assumed in many of the ML FFs mentioned earlier. Mapping atomic fingerprints to purely atomic properties can thus lead to powerful and accurate prescriptions. One such methodology called AGNI has been shown work well for a variety of elements (Al, C, Pt, etc.) for a variety of mechanical and thermal properties \cite{botu2017machine, huan2019iterative, chapman2020machine, batra2019general}.

\begin{figure}
	\centering 
	\includegraphics[width=0.75\textwidth]{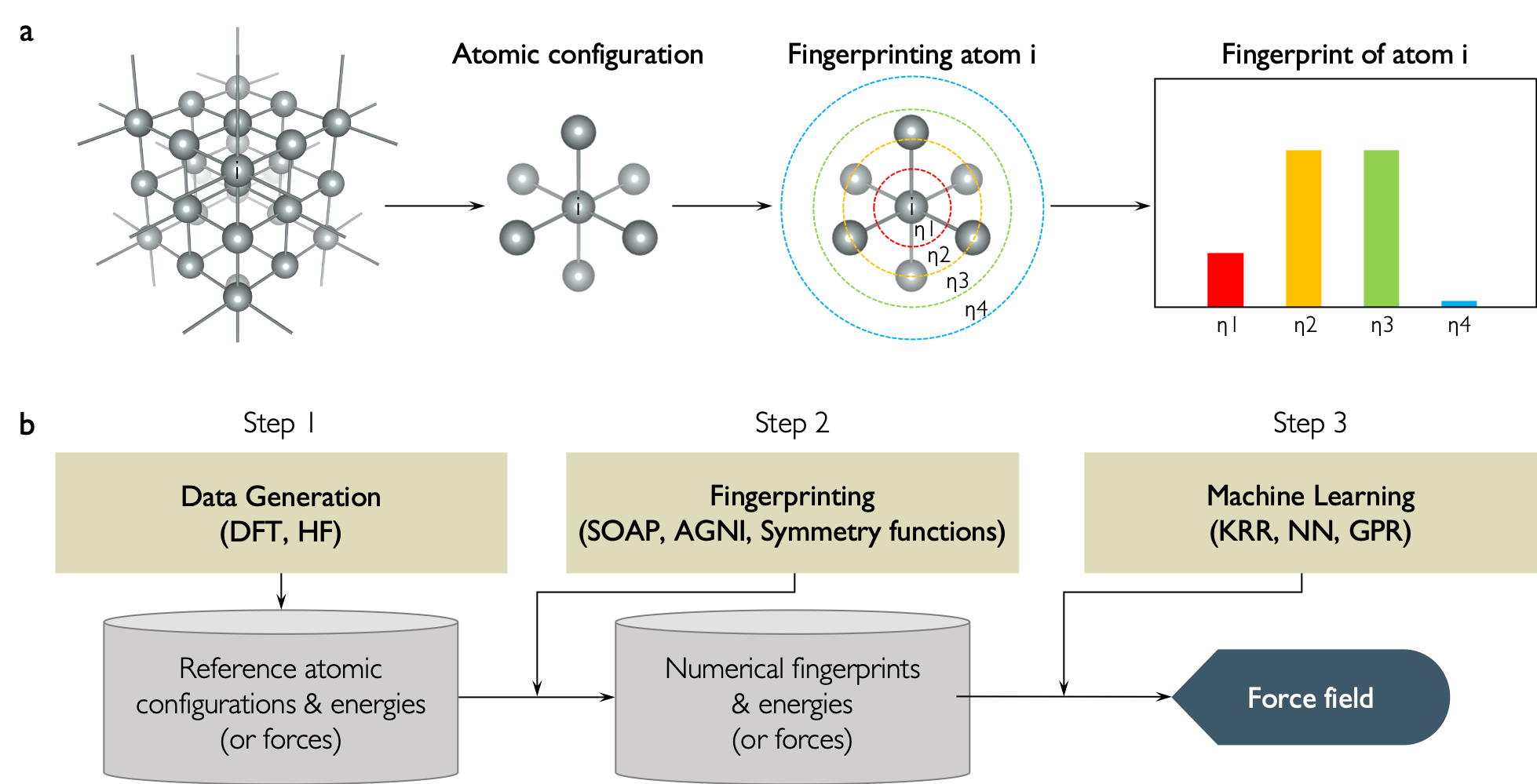}
	\caption{Building FF using machine learned functional form. Adapted from \cite{ramprasad2017machine}.}
	\label{fig:figure2}
\end{figure}

Another promising development in this area has been the use of graph networks \cite{isayev2017universal, xie2018crystal, chen2019graph, gilmer2017neural}, which by construction overcome the limitations of manually constructed atomic fingerprints (or feature vectors). A graph representation of molecules or crystals consists of feature vectors capturing information about all the constituting atoms (nodes/vertex), and bonds or atomic distances (edge). This graph representation can be fed into a ML model (neural networks) and can be mapped to material property of interest. So far this methodology has not been used for FF development, but the unprecedented level accuracy achieved by such models on the Materials Project \cite{jain2013commentary} and the QM9 databases \cite{ramakrishnan2014quantum} to predict diverse properties, such as energy, polarization, vibrational frequency, band gap, bulk modulus, etc., offers great promise.

Previously, we discussed how GA can be used to effectively find parameters of a FF with known functional form. However, GA can also be utilized to identify new functional forms themselves using symbolic regression \cite{sun2019data}. The key to this approach is to search within a physically meaningful hypothesis space. Analysis of interatomic potentials derived over the past several decades from physical principles, reveals that only a few fundamental operations are necessary to recover functional form of these classical potentials. These operations include addition, subtraction, multiplication, division, and power operators; constant values and distances between atoms; and an operator that performs a sum over functions of distances between a given atom and all neighbors within a given cutoff radius. Thus, if the GA is used to search within a combinatorial space of these physically meaningful operations, new FF functional forms can be learned that better fit the available quantum mechanical or empirical data. Symbolic regression is another crucial piece of this methodology which transforms a given functional form into a tree structure, as shown in Figure 3c, to make it amenable to basic GA crossover and mutation operations. Mueller and co-workers \cite{hernandez2019fast} successfully used this scheme to not only recover the exact form of famous Lennard-Jones and Sutton-Chen (SC) EAM potential, but also build ML potentials for Cu using DFT training data, with an exemplary potential being $E_i = \sum (r^{10.21-5.47r} -0.21^{r}) f(r) + 0.97 \big(\sum 0.33^{r}f(r) \big)^{-1}$. 

\section{Iterative Improvement of FF using Active and Transfer Learning}
A key limitation of any FF, whether traditional or ML based, is that they may not work for applications (simulations) involving configurational domains that are different from their training domain \cite{bianchini2016modelling}. As such the transferability of the potential function or force-field is an important measure of its predictive power. For example, FFs are known to be unsuitable for simulations far from equilibrium, as they are mostly trained to reproduce equilibrium properties and configurations. Likewise, neural network based potentials are interpolative and as such tend to fail miserably when encountering scenarios outside of their training set \cite{pun2019physically, chen2019machine}. Despite these challenges, the ML approach does provide an opportunity to systematically improve versatility and transferability of FFs over the traditional pre-defined functional forms.

The interpolative nature of the ML FF, especially those based on NN, demands a large number of training data to ensure that all possible physical scenarios are adequately captured. Their training, therefore, often requires large quantities ($\sim$10$^4$ or greater) of data to ensure that the model adequately samples the energy landscape both near and far-from-equilibrium. A highly desirable goal is to minimize the number of training examples, especially if the underlying reference model is first-principles based and hence computationally expensive. This has led to the emergence of ML techniques such as `active learning' and `transfer learning' that focus on efficient sampling of the training data. The overarching goal is to reduce the number of training data to the absolute minimum without compromising on the adequate representation of various areas in a potential energy surface.

There have been many recent studies aimed at addressing the sampling challenge in generation of training data. Smith et al. \cite{smith2018less} employed an active learning (AL) strategy based on the Query by Committee (QBC) scheme. QBC uses the disagreement between an ensemble of ML potentials to infer the reliability of the ensemble's prediction. QBC allowed for automatic sampling of the regions of chemical space where the potential energy was not accurately described by the ML potential. Their AL approach was validated on a test set consisting of a diverse set of organic molecules and their results showed that one requires only 10\% to 25\% of the data to accurately represent the chemical space of these molecules. Similarly, Zhang et al. \cite{zhang2019active} introduced an AL scheme (deep potential generator (DP-GEN)) that constructs ML models for simulating materials at the molecular scale. Their procedure involve exploration, generation of accurate reference data, and training. They used Al and Al-Mg as representative cases and showed that ML models can be trained with minimum number of reference data. Another interesting body of work in this direction is on-the-fly learning to accelerate AIMD. These methods rely on building FFs that not only provide point estimates of energy and atomic forces, but additionally outputs associated prediction uncertainty. \textcolor{black}{Ceriotti and co-workers have developed an inexpensive yet reliable way of estimating this uncertainty associated with ML model predictions of atomic and molecular properties. Their scheme is based on resampling wherein they generate multiple models based on subsampling of the same training data. They benchmark the accuracy of the uncertainty prediction by maximum likelihood estimation which in turn can correct for correlations between resampled models and improve performance of the uncertainty estimation via a cross-validation procedure \cite{musil2019fast}}. By tracking model uncertainty during the MD simulation, a call to the high-fidelity (but expensive) DFT calculation can made when the system drifts to a configuration where the model is uncertain in the energy/force prediction beyond a certain threshold \cite{li2015molecular, jinnouchi2019phase, jinnouchi2019fly}. Vandermause et al. \cite{vandermause2019fly} sampled structures on-the-fly from AIMD and used an adaptive Bayesian inference method to automate the training of low-dimensional multiple element interatomic force fields. Their AL framework uses internal uncertainty of a GPR model to decide acceptance of model prediction or the need to augment training data. In all of the above studies, the overarching aim is to minimize the \emph{ab-initio} training data required to describe the potential energy surface.

In another exemplary work by Huan and co-workers \cite{huan2019iterative}, ``failed" data augmentation was used to systematically improve the domain of applications (and configurations) that can be handled using the AGNI ML FF. They systematically tested their ML FFs on new applications (stress/strain analysis, stacking fault energetics, and melting simulations), identified configurations where the present version of FF failed, appended those configurations to the original training data, and retrained the ML model on the updated training data, thereby increasing the domain of applicability of their FFs. Another crucial information that was transferred from the previous FF version during the retraining process was all the associated ML parameters, e.g., the fingerprint parameters, the learning algorithm, and the previously selected training data. This helped transfer the current performance/knowledge of the FF to the next generation which was then trained using the augmented dataset. More recently, Loeffler et al. \cite{loeffler2020active} have developed an active learning approach that iteratively trains an ANN model to faithfully reproduce the coarse-grained potential energy surface of water clusters. They showed that one can train a NN potential with only 426 total structures in its training data. Their AL workflow initiates with a sparse training data set and is continually updated \textit{via} a Nested Ensemble Monte Carlo scheme that sparsely queries the energy landscape and tests the network performance.  The network is retrained with an updated training set that includes failed configurations/energies from previous iteration until convergence is attained. Such a trained network adequately reproduces the energies (within mean absolute error (MAE) of 2 meV/molecule) and forces (MAE 40 meV/Å) compared to the reference model. Such studies have laid the groundwork on for workflows that allow on-demand generation of training data.







\section{Future Directions and Perspective}

With the recent advances in computing power, AI, ML and big data analytics are emerging as powerful tools for materials modeling. The flexibility, accuracy and computational efficiency of ML models can accelerate materials discovery and design and can also be leveraged for advanced nanoscale materials characterization. The ML models are primarily geared towards addressing one of the most critical aspects of MD i.e. its predictive power which hinges strongly on the nature of the interatomic potential used to describe the atomistic interactions in the system. The various examples presented in this perspective article highlight the advantages of using ML and data driven approaches for training new models. There is, however, still a lot of room for improvement and we believe that both supervised and unsupervised ML methods can play a key role towards the development of next-generation atomistic, molecular or coarse-grained models. Some of these areas are briefly discussed below.

MD with pre-defined functional forms continue to remain attractive owing to their computational efficiency. \textcolor{black}{One advantage of pre-defined force-fields is that they are usually based on physical models that attempt to reliably capture atomistic interactions. This physical basis imparts predictability and transferability for such models. However, most functional forms are restricted to first order treatments of the physical models and ignore higher level terms owing to the increase in computational cost. This is especially true for polarizable or reactive systems, where the treatment of electrostatics still relies on point charges on atoms. To adequately describe a molecular dipole moment and its electrostatic field, one needs to include dipole moments of the atomic charge distributions. While this was established decades ago by Stone AJ \cite{stone1981distributed}, such terms are typically not included owing to their computational complexity, and therefore the accompanying polarization effects  directly described. Likewise, the charge transfer dynamics in the case of reactive systems is still described by methods such as electronegativity equalization method (EEM) \cite{mortier1986electronegativity} and charge equilibration (Qeq) \cite{rappe1991charge}. The charge calculations represent the most compute intensive part of reactive potential models. These charge transfer and polarizability effects can have strong significance for organic systems, and for glasses and ceramics, especially at the surfaces and interfaces. The advances in ML and the automated training processes should enable us to revisit and efficiently describe effects arising from higher order terms to adequately capture electrostatic contributions in polarizable and reactive systems.}

While pre-defined functional forms are efficient, the choice of a pre-defined function limits their ability to capture the underlying chemistry and physics being modeled. Despite a lot of advances in sampling and generation of training data as well as improvements in global optimization algorithms, there is always going to be a ceiling limit for pre-defined functional forms. Flexibility in the functional form is one of the critical challenges. The re-emergence of neural network models in the post-AI winter era is aimed at providing more flexibility to users interested in modeling complex reactive systems that involve multiple bonding characteristics (e.g. metallic, covalent and ionic). However, the advantage offered by such flexible ML potentials can also become their limitation; when such ML potentials are used outside their domain of applicability (far from training data) they result in arbitrary unphysical behavior. Thus, more research on constraining these models to basic physical relations, such as the work on physics informed neural networks \cite{pun2019physically}, will be highly useful.

All ML frameworks require carefully curation and sampling of training data. In the context of materials models, one should ensure that the training data used is sufficiently diverse (for instance, training set that consists of configurations spanning broad range of energies from near equilibrium to highly non-equilibrium). For the model to be robust, one has to ensure ample representation of the different parts of the potential energy surface i.e. both near equilibrium and far-from-equilibrium configurations in their training data. The availability of first-principles training datasets via Materials Project and other thermodynamic databases puts us in an enviable position to incorporate robustness in the ML developed models. However, it still is not properly understood what is the minimum diversity and size of training dataset required to build robust FFs. On the other hand, advances in workflows have also allowed for direct interfacing with \textcolor{black}{electronic structure codes (e.g. VASP)} allowing users to generate training data on-the-fly. This is an important emerging area and recent studies, especially on active and transfer learning strategies, highlight the importance of sampling and incorporation of diversity in training data.

In this respect, one of the significant departures in MD potential development compared to conventional fitting is the development of automated training workflows that explicitly include temperature dependent properties in its objective function. Traditionally, one would define objective functions that would primarily include the static properties such as configurational energies, cohesive energies, lattice constants to name a few. Such models, however, lack predictive power to capture dynamical and transport properties. Our recent work has introduced workflows that allow temperature dependent properties derived from on-the-fly molecular dynamics to be included as part of the training procedure. The ability to directly train MD potentials that capture transport properties and other temperature/pressure dependent properties is of tremendous significance to capture dynamical phenomena near phase boundaries. \textcolor{black}{In this regard, Chan et al. trained a bond order potential for WSe$_2$ which was used to study thermal transport behavior of diverse nanostructures at different temperatures using non-equilibrium MD simulation\cite{chan2019machineb}. The potential was fit to not only first principles data, but the temperature dependent stability of WSe$_2$ structures was also among the fitting criteria. The idea is to run on-the-fly MD simulations during the potential fitting process to assess the stability of various structures at different target temperatures\cite{chan2019machinea, chan2019machineb}.}

\textcolor{black}{It should be noted that MD simulations with classical Newtonian mechanics ignore quantum zero-point energy contributions to structural fluctuations. The quantization of the energy of the vibration modes cannot be accounted for by using standard MD because it is based on classical statistics. In this regard, Dammak et al. \cite{dammak2009quantum} introduced a quantum thermal bath (QTB) that accounts for quantum statistics while using standard MD. The basic idea of the QTB is to use a Langevin-type approach. Barrat et al. implemented QTB in a flexible manner and at a low computational cost by synthesizing the corresponding noise `on the fly' \cite{binder2011monte}. These corrections are especially important when evaluating thermal and other transport properties. We note that the effect for metals (especially transition elements) is generally small, but it becomes more important for glasses and ceramics, and very significant for organics. Thus, depending on the material type, such corrections should not be ignored and become important when one includes temperature dependent quantities in future training procedures.}

Multi-fidelity scale bridging from the electronic structure to coarse-grained scales emphasizes a greater need to estimate the extent and nature of error propagation from the high-fidelity first-principles scale to coarse-grained mesoscopic scale. With the increasing use of higher-level theory (CCSD or QMC) to generate the training data and reduce the error necessitates quantification of uncertainties in the predictions at various scales. In this regard, Functional Uncertainty Quantification method (FunUQ) \cite{reeve2017error} to assess model form errors, such as interatomic potentials for MD simulations represents an important future direction. Cross-validation, sensitivity analysis and uncertainty quantification will be critical to test the robustness of their developed models. Preliminary work on the use of uncertainty quantification has also lead to autonomous MD workflows such as FLARE \cite{vandermause2019fly} that provide a merger between high-fidelity quantum mechanical and low-fidelity machine learning potentials.

\textcolor{black}{We note that there are limitations to DFT---for instance, vibrational frequencies/phonon spectra calculated by DFT can have errors of $\sim$10\% and this could represent a source of error in the training set. One therefore requires some sort of calibration to bring the calculated data in closer agreement with experiments (reality). This could be achieved by performing more higher level quantum calculations (CCSD or QMC) to capture the errors in DFT and establish some scaling relationship or via transfer learning where the weights of a neural network are first trained using DFT and then correctly by performing a limited number of higher level DFT calculations.}

ML global optimization algorithms and multi-stage optimization strategies clearly represent a major advance over local optimization procedures that have been traditionally employed for optimizing parameters to describe the potential energy surface. Evolutionary optimization procedures such as genetic algorithms remain attractive but one can easily envision the use of other emerging AI techniques such as Monte Carlo Tree Search (MCTS) \cite{silver2016mastering} to derive globally optimal solutions. The use of decision trees and dynamic learning to navigate more effectively through the potential energy landscape is expected to address some of the bottlenecks with the sluggishness of global optimization schemes.

A key aspect of optimization is to appropriately define objective functions that go beyond the simple sum of square difference. The selection of the appropriate weights poses major problems for optimization that focus on single objective functions. \textcolor{black}{The weight selection in a single objective function is typically dependent on the expertise of the force-field developer. The goodness of fit (or the predictive capability of the force-field) is measured by a weighted sum of errors in predictions of a set of target observables. The general rule of thumb is that the weights should be selected such that the contribution to the overall error for each of the individual properties in the objective function is approximately the same. Nonetheless, the weight selection is quite challenging and typically, the force-fields developed tend to do well in a few properties and rather poorly in others---even though the overall objective function has reached a minimum. To circumvent the issues with weight selection, there have been developments in multi-objective strategies where each of the objectives is defined separately. One such strategy is based on Pareto optimization, where one retains a set of optimal solutions and the user can choose a solution depending on the properties of interest. A Pareto-optimal solution is one in which a small variation of parameters that leads to an improvement in any of the objectives will lead to at least one of the other objectives being less optimal; there can be curves or surfaces in the parameter space corresponding to such solutions and these are termed Pareto fronts. Such multi-objective evolutionary approaches are especially helpful if one is dealing with conflicting properties (e.g. elastic constants vs. defect formation energies). Other multi-objective approaches that have recently emerged are those that treat a hierarchy of objectives during an evolutionary search\cite{chan2019machinea}. In this case, the quality of a proposed parameter set is evaluated based on a hierarchical objective function. In this evolutionary scheme, we truncate the evaluation of any parameter set which leads to large errors in hierarchical property classes and assign it a penalty depending on which class it fails at. This approach aids in an accelerated evolutionary search by (i) efficiently sampling the parameter landscape within a given generation, and (ii) overcoming the limitation of assigning arbitrary weights within a single objective thereby ensuring that all the properties (static or dynamic) are equally well described. Such an approach has been used recently to train a coarse-grained model for water. This model is able to successfully capture the various thermodynamic anomalies of water and outperforms most other existing atomistic and coarse-grained models in their predictive power\cite{chan2019machinea}.} Multi-objective optimization such as \textcolor{black}{Pareto} optimization and the recent success of HOGA (hierarchical objective genetic algorithm) \cite{chan2019machinea} to find optimal solutions that represent a compromise between the various desired objectives is a step in the right direction. One needs to exercise caution in defining the objective appropriately otherwise even the best global optimization strategy may not yield the best results.

Finally, while a lot of work has focused on length-scale bridging, timescale challenges remain unaddressed. All-atom simulations typically employ 0.1-1 femtosecond timesteps and as such are suited for modeling phenomena in the nanosecond timescales. Even with exascale computing, one would primarily gain in terms of spatial scales that can be accessed. The clock-speeds and bandwidths are not going to change significantly and timescale challenges will remain. \textcolor{black}{ Coarse-graining allows for larger time-step during MD integration and thereby explore longer time-scales. In particular, coarse-grained models  typically allow 10-50 femtosecond timesteps and represent one route to address this timescale challenge albeit typically at the cost of accuracy. The use of coarse-grained models along with advanced sampling techniques is becoming very popular among the community for exploring longer time-scales, exploring phase-transitions such as Lower/Upper Critical Solution Temperature LCST/UCST in polymeric solutions and brushes. In particular, the MARTINI coarse grain model \cite{marrink2007martini} has gained in popularity and is increasingly being used to understand drug uptakes and transport across lipid bilayer membrane and membrane based dynamics. The future looks promising with the advent of quantum computing and the use of physics-based polarizable models coupled with path integral calculations for accurately determining the free energy of binding of drugs and also exploring of time-scale where rare events such as allosteric modulation, molecular recognition, protein folding and self-assembly. To explore such time-scales, chemists are resorting to polarizable united-atom and coarse-grained models that can more adequately treat electrostatics.}

\textcolor{black}{One of the major challenges with coarse-grained models is that the gain in efficiency comes at the cost of accuracy. This is particularly true for water, which is one of the most important solvent used in soft-matter simulations of polymers, proteins and other biomolecules. Traditionally, models to describe water range from fixed charge dispersion/repulsion 12-6 model, Buckingham potential, Stockmayer potential, ST2, force fields due to fluctuating charge, and the inclusion of van-der Waal parameters for explicit hydrogens have been explored in the past 80 years \cite{onufriev2018water}. Further refinements include the introduction of sites on the HOH angle bisector, multiple sites $>$3 for water based on the TIPnP framework. The target data for water includes the phase diagram, ice-formation abilities, structural properties (radial distribution functions), transport properties including diffusion, thermal conductivity and viscosity not to forget the anomalous density variation with temperature of dense water. ML strategies can allow for development of coarse-grained surrogates that address the timescale challenges without sacrificing accuracy. ML CG water models developed by Chan et al. \cite{chan2019machinea} that allow for on-the-fly sampling of temperature dependent properties have succeeded in capturing the various thermodynamic properties and anomalies of water at fraction of the computational cost of its atomistic counterpart.} Such coarser surrogate model development augurs well for capturing hitherto unexplored dynamical phenomena in the mesoscopic regime.

\bibliographystyle{unsrt}

\begin{thebibliography}{100}

\bibitem{rahman1964correlations}
Aneesur Rahman.
\newblock Correlations in the motion of atoms in liquid argon.
\newblock {\em Phys. Rev.}, 136(2A):A405, 1964.

\bibitem{de2016role}
Marco De~Vivo, Matteo Masetti, Giovanni Bottegoni, and Andrea Cavalli.
\newblock Role of molecular dynamics and related methods in drug discovery.
\newblock {\em J. Med. Chem.}, 59(9):4035--4061, 2016.

\bibitem{liu2018molecular}
Xuewei Liu, Danfeng Shi, Shuangyan Zhou, Hongli Liu, Huanxiang Liu, and Xiaojun
  Yao.
\newblock Molecular dynamics simulations and novel drug discovery.
\newblock {\em Expert Opin Drug Dis.}, 13(1):23--37, 2018.

\bibitem{frederix2015exploring}
Pim~WJM Frederix, Gary~G Scott, Yousef~M Abul-Haija, Daniela Kalafatovic,
  Charalampos~G Pappas, Nadeem Javid, Neil~T Hunt, Rein~V Ulijn, and Tell
  Tuttle.
\newblock Exploring the sequence space for (tri-) peptide self-assembly to
  design and discover new hydrogels.
\newblock {\em Nat. Chem.}, 7(1):30, 2015.

\bibitem{hollingsworth2018molecular}
Scott~A Hollingsworth and Ron~O Dror.
\newblock Molecular dynamics simulation for all.
\newblock {\em Neuron}, 99(6):1129--1143, 2018.

\bibitem{sresht2015liquid}
Vishnu Sresht, Agilio~AH Padua, and Daniel Blankschtein.
\newblock Liquid-phase exfoliation of phosphorene: design rules from molecular
  dynamics simulations.
\newblock {\em ACS Nano}, 9(8):8255--8268, 2015.

\bibitem{singh2015computational}
Arunima~K Singh, Kiran Mathew, Houlong~L Zhuang, and Richard~G Hennig.
\newblock Computational screening of {2D} materials for photocatalysis.
\newblock {\em J. Phys. Chem. Lett.}, 6(6):1087--1098, 2015.

\bibitem{belmares2004hildebrand}
M~Belmares, M~Blanco, WA~Goddard~Iii, RB~Ross, G~Caldwell, S-H Chou, J~Pham,
  PM~Olofson, and Cristina Thomas.
\newblock Hildebrand and hansen solubility parameters from molecular dynamics
  with applications to electronic nose polymer sensors.
\newblock {\em J. Comput. Chem.}, 25(15):1814--1826, 2004.

\bibitem{bagri2011thermal}
Akbar Bagri, Sang-Pil Kim, Rodney~S Ruoff, and Vivek~B Shenoy.
\newblock Thermal transport across twin grain boundaries in polycrystalline
  graphene from nonequilibrium molecular dynamics simulations.
\newblock {\em Nano Lett.}, 11(9):3917--3921, 2011.

\bibitem{skoulidas2005self}
Anastasios~I Skoulidas and David~S Sholl.
\newblock Self-diffusion and transport diffusion of light gases in
  metal-organic framework materials assessed using molecular dynamics
  simulations.
\newblock {\em J. Phys. Chem. B}, 109(33):15760--15768, 2005.

\bibitem{luo2012enhancement}
Tengfei Luo and John~R Lloyd.
\newblock Enhancement of thermal energy transport across graphene/graphite and
  polymer interfaces: a molecular dynamics study.
\newblock {\em Adv. Func. Mater.}, 22(12):2495--2502, 2012.

\bibitem{hura2003water}
Greg Hura, Daniela Russo, Robert~M Glaeser, Teresa Head-Gordon, Matthias Krack,
  and Michele Parrinello.
\newblock Water structure as a function of temperature from {X-ray} scattering
  experiments and ab initio molecular dynamics.
\newblock {\em Phys. Chem. Chem. Phys.}, 5(10):1981--1991, 2003.

\bibitem{chen2018thermal}
Yachao Chen, Sukriti Manna, Cristian~V Ciobanu, and Ivar~E Reimanis.
\newblock Thermal regimes of {Li}-ion conductivity in $\beta$-eucryptite.
\newblock {\em J. Am. Ceram. Soc.}, 101(1):347--355, 2018.

\bibitem{hafner2008ab}
J{\"u}rgen Hafner.
\newblock Ab-initio simulations of materials using {VASP}: Density-functional
  theory and beyond.
\newblock {\em J. Comput. Chem.}, 29(13):2044--2078, 2008.
\newblock \url{https://www.vasp.at}.

\bibitem{plimpton1993fast}
Steve Plimpton.
\newblock Fast parallel algorithms for short-range molecular dynamics.
\newblock Technical report, Sandia National Labs., Albuquerque, NM (United
  States), 1993.
\newblock \url{http://lammps.sandia.gov}.

\bibitem{larsen2017atomic}
Ask~Hjorth Larsen, Jens~J{\o}rgen Mortensen, Jakob Blomqvist, Ivano~E Castelli,
  Rune Christensen, Marcin Du{\l}ak, Jesper Friis, Michael~N Groves, Bj{\o}rk
  Hammer, Cory Hargus, et~al.
\newblock The atomic simulation environment—a python library for working with
  atoms.
\newblock {\em J. Phys-Condens. Mat.}, 29(27):273002, 2017.
\newblock \url{https://wiki.fysik.dtu.dk/ase}.

\bibitem{phillips2005scalable}
James~C Phillips, Rosemary Braun, Wei Wang, James Gumbart, Emad Tajkhorshid,
  Elizabeth Villa, Christophe Chipot, Robert~D Skeel, Laxmikant Kale, and Klaus
  Schulten.
\newblock Scalable molecular dynamics with namd.
\newblock {\em J. Comput. Chem.}, 26(16):1781--1802, 2005.
\newblock \url{https://www.ks.uiuc.edu/Research/namd}.

\bibitem{brooks2009charmm}
Bernard~R Brooks, Charles~L Brooks~III, Alexander~D Mackerell~Jr, Lennart
  Nilsson, Robert~J Petrella, Beno{\^\i}t Roux, Youngdo Won, Georgios
  Archontis, Christian Bartels, Stefan Boresch, et~al.
\newblock {CHARMM}: the biomolecular simulation program.
\newblock {\em J. Comput. Chem.}, 30(10):1545--1614, 2009.
\newblock \url{https://www.charmm.org/charmm}.

\bibitem{berendsen1995gromacs}
Herman~JC Berendsen, David van~der Spoel, and Rudi van Drunen.
\newblock {GROMACS}: a message-passing parallel molecular dynamics
  implementation.
\newblock {\em Comput. Phys. Commun.}, 91(1-3):43--56, 1995.
\newblock \url{http://www.gromacs.org}.

\bibitem{hohenberg1964inhomogeneous}
Pierre Hohenberg and Walter Kohn.
\newblock Inhomogeneous electron gas.
\newblock {\em Phys. Rev.}, 136(3B):B864, 1964.

\bibitem{kohn1965self}
Walter Kohn and Lu~Jeu Sham.
\newblock Self-consistent equations including exchange and correlation effects.
\newblock {\em Phys. Rev.}, 140(4A):A1133, 1965.

\bibitem{car1985unified}
Richard Car and Mark Parrinello.
\newblock Unified approach for molecular dynamics and density-functional
  theory.
\newblock {\em Phys. Rev. Lett.}, 55(22):2471, 1985.

\bibitem{chan2019machinec}
Henry Chan, Badri Narayanan, Mathew~J Cherukara, Fatih~G Sen, Kiran Sasikumar,
  Stephen~K Gray, Maria~KY Chan, and Subramanian~KRS Sankaranarayanan.
\newblock Machine learning classical interatomic potentials for molecular
  dynamics from first-principles training data.
\newblock {\em J. Phys. Chem. C}, 123(12):6941--6957, 2019.

\bibitem{kim2018polymer}
Chiho Kim, Anand Chandrasekaran, Tran~Doan Huan, Deya Das, and Rampi Ramprasad.
\newblock Polymer genome: a data-powered polymer informatics platform for
  property predictions.
\newblock {\em J. Phys. Chem. C}, 122(31):17575--17585, 2018.

\bibitem{mueller2016machine}
Tim Mueller, Aaron~Gilad Kusne, and Rampi Ramprasad.
\newblock Machine learning in materials science: Recent progress and emerging
  applications.
\newblock {\em Rev. Comput. Chem.}, 29:186--273, 2016.

\bibitem{butler2018machine}
Keith~T Butler, Daniel~W Davies, Hugh Cartwright, Olexandr Isayev, and Aron
  Walsh.
\newblock Machine learning for molecular and materials science.
\newblock {\em Nature}, 559(7715):547--555, 2018.

\bibitem{ward2017atomistic}
Logan Ward and Chris Wolverton.
\newblock Atomistic calculations and materials informatics: A review.
\newblock {\em Curr. Opin. Solid State Mater. Sci.}, 21(3):167--176, 2017.

\bibitem{chandrasekaran2019solving}
Anand Chandrasekaran, Deepak Kamal, Rohit Batra, Chiho Kim, Lihua Chen, and
  Rampi Ramprasad.
\newblock Solving the electronic structure problem with machine learning.
\newblock {\em Npj Comput. Mater.}, 5(1):1--7, 2019.

\bibitem{pilania2017multi}
Ghanshyam Pilania, James~E Gubernatis, and Turab Lookman.
\newblock Multi-fidelity machine learning models for accurate bandgap
  predictions of solids.
\newblock {\em Comput. Mater. Sci.}, 129:156--163, 2017.

\bibitem{jones1924determination}
John~Edward Jones.
\newblock On the determination of molecular fields.—{II}. {From} the equation
  of state of a gas.
\newblock {\em Proc. R. Soc. of Lon. Ser-A}, 106(738):463--477, 1924.

\bibitem{chenoweth2008reaxff}
Kimberly Chenoweth, Adri~CT Van~Duin, and William~A Goddard.
\newblock {ReaxFF} reactive force field for molecular dynamics simulations of
  hydrocarbon oxidation.
\newblock {\em J. Phys. Chem. A}, 112(5):1040--1053, 2008.

\bibitem{shan2010charge}
Tzu-Ray Shan, Bryce~D Devine, Travis~W Kemper, Susan~B Sinnott, Simon~R
  Phillpot, et~al.
\newblock Charge-optimized many-body potential for the hafnium/hafnium oxide
  system.
\newblock {\em Phys. Rev. B}, 81(12):125328, 2010.

\bibitem{mockus2012bayesian}
Jonas Mockus.
\newblock {\em Bayesian approach to global optimization: theory and
  applications}, volume~37.
\newblock Springer Science \& Business Media, 2012.

\bibitem{mitchell1998introduction}
Melanie Mitchell.
\newblock {\em An introduction to genetic algorithms}.
\newblock MIT press, 1998.

\bibitem{rasmussen2003gaussian}
Carl~Edward Rasmussen.
\newblock Gaussian processes in machine learning.
\newblock In {\em Summer School on Machine Learning}, pages 63--71. Springer,
  2003.

\bibitem{liu2019parameterization}
Han Liu, Zipeng Fu, Yipeng Li, Nazreen Farina~Ahmad Sabri, and Mathieu Bauchy.
\newblock Parameterization of empirical forcefields for glassy silica using
  machine learning.
\newblock {\em MRS Commun.}, 9(2):593--599, 2019.

\bibitem{movckus1975bayesian}
Jonas Mo{\v{c}}kus.
\newblock On bayesian methods for seeking the extremum.
\newblock In {\em Optimization techniques IFIP technical conference}, pages
  400--404. Springer, 1975.

\bibitem{takahashi1998convergence}
Yoshikane Takahashi.
\newblock Convergence of simple genetic algorithms for the two-bit problem.
\newblock {\em BioSystems}, 46(3):235--282, 1998.

\bibitem{corne2018evolutionary}
David~W Corne and Michael~A Lones.
\newblock Evolutionary algorithms.
\newblock {\em arXiv preprint arXiv:1805.11014}, 2018.

\bibitem{larsson2013global}
Henrik~R Larsson, Adri~CT van Duin, and Bernd Hartke.
\newblock Global optimization of parameters in the reactive force field
  {ReaxFF} for {SiOH}.
\newblock {\em J. Comput. Chem.}, 34(25):2178--2189, 2013.

\bibitem{chan2019machinea}
Henry Chan, Mathew~J Cherukara, Badri Narayanan, Troy~D Loeffler, Chris
  Benmore, Stephen~K Gray, and Subramanian~KRS Sankaranarayanan.
\newblock Machine learning coarse grained models for water.
\newblock {\em Nat. Commun.}, 10(1):1--14, 2019.

\bibitem{chan2019machineb}
Henry Chan, Kiran Sasikumar, Srilok Srinivasan, Mathew Cherukara, Badri
  Narayanan, and Subramanian~KRS Sankaranarayanan.
\newblock Machine learning a bond order potential model to study thermal
  transport in {WSe2} nanostructures.
\newblock {\em Nanoscale}, 11(21):10381--10392, 2019.

\bibitem{loeffler2019teaching}
Troy~D Loeffler, Henry Chan, Kiran Sasikumar, Badri Narayanan, Mathew~J
  Cherukara, Stephen Gray, and Subramanian~KRS Sankaranarayanan.
\newblock Teaching an old dog new tricks: Machine learning an improved tip3p
  potential model for liquid--vapor phase phenomena.
\newblock {\em J. Phys. Chem. C}, 123(36):22643--22655, 2019.

\bibitem{narayanan2016multi}
Badri Narayanan, Kiran Sasikumar, Zhi-Gang Mei, Alper Kinaci, Fatih~G. Sen,
  Michael~J. Davis, Stephen~K. Gray, Maria K.~Y. Chan, and Subramanian K. R.~S.
  Sankaranarayanan.
\newblock Development of a modified embedded atom force field for zirconium
  nitride using multi-objective evolutionary optimization.
\newblock {\em The Journal of Physical Chemistry C}, 120(31):17475--17483,
  2016.

\bibitem{narayanan2016gold}
Badri Narayanan, Alper Kinaci, Fatih~G. Sen, Michael~J. Davis, Stephen~K. Gray,
  Maria K.~Y. Chan, and Subramanian K. R.~S. Sankaranarayanan.
\newblock Describing the diverse geometries of gold from nanoclusters to
  bulk—a first-principles-based hybrid bond-order potential.
\newblock {\em The Journal of Physical Chemistry C}, 120(25):13787--13800,
  2016.

\bibitem{fatih2014iridium}
F.~G. Sen, A.~Kinaci, B.~Narayanan, S.~K. Gray, M.~J. Davis, S.~K. R.~S.
  Sankaranarayanan, and M.~K.~Y. Chan.
\newblock Towards accurate prediction of catalytic activity in {IrO2}
  nanoclusters via first principles-based variable charge force field.
\newblock {\em J. Mater. Chem. A}, 3:18970--18982, 2015.

\bibitem{cherukara2016stanene}
Mathew~J. Cherukara, Badri Narayanan, Alper Kinaci, Kiran Sasikumar, Stephen~K.
  Gray, Maria~K.Y. Chan, and Subramanian K. R.~S. Sankaranarayanan.
\newblock Ab initio-based bond order potential to investigate low thermal
  conductivity of stanene nanostructures.
\newblock {\em The Journal of Physical Chemistry Letters}, 7(19):3752--3759,
  2016.
\newblock PMID: 27569053.

\bibitem{silver2016mastering}
David Silver, Aja Huang, Chris~J Maddison, Arthur Guez, Laurent Sifre, George
  Van Den~Driessche, Julian Schrittwieser, Ioannis Antonoglou, Veda
  Panneershelvam, Marc Lanctot, et~al.
\newblock Mastering the game of {Go} with deep neural networks and tree search.
\newblock {\em Nature}, 529(7587):484, 2016.

\bibitem{wilde2011swift}
Michael Wilde, Mihael Hategan, Justin~M Wozniak, Ben Clifford, Daniel~S Katz,
  and Ian Foster.
\newblock Swift: A language for distributed parallel scripting.
\newblock {\em Parallel Computing}, 37(9):633--652, 2011.

\bibitem{dreher2017decaf}
Matthieu Dreher and Tom Peterka.
\newblock Decaf: Decoupled dataflows for in situ high-performance workflows.
\newblock Technical report, Argonne National Lab.(ANL), Argonne, IL (United
  States), 2017.

\bibitem{finnis2007bond}
MW~Finnis.
\newblock Bond-order potentials through the ages.
\newblock {\em Prog. Mater. Sci.}, 52(2-3):133--153, 2007.

\bibitem{botu2017machine}
Venkatesh Botu, Rohit Batra, James Chapman, and Rampi Ramprasad.
\newblock Machine learning force fields: construction, validation, and outlook.
\newblock {\em J. Phys. Chem. C}, 121(1):511--522, 2017.

\bibitem{huan2017universal}
Tran~Doan Huan, Rohit Batra, James Chapman, Sridevi Krishnan, Lihua Chen, and
  Rampi Ramprasad.
\newblock A universal strategy for the creation of machine learning-based
  atomistic force fields.
\newblock {\em Npj Comput. Mater.}, 3(1):1--8, 2017.

\bibitem{kuritz2018size}
Natalia Kuritz, Goren Gordon, and Amir Natan.
\newblock Size and temperature transferability of direct and local deep neural
  networks for atomic forces.
\newblock {\em Phys. Rev. B}, 98(9):094109, 2018.

\bibitem{rowe2018development}
Patrick Rowe, G{\'a}bor Cs{\'a}nyi, Dario Alf{\`e}, and Angelos Michaelides.
\newblock Development of a machine learning potential for graphene.
\newblock {\em Phys. Rev. B}, 97(5):054303, 2018.

\bibitem{podryabinkin2019accelerating}
Evgeny~V Podryabinkin, Evgeny~V Tikhonov, Alexander~V Shapeev, and Artem~R
  Oganov.
\newblock Accelerating crystal structure prediction by machine-learning
  interatomic potentials with active learning.
\newblock {\em Phys. Rev. B}, 99(6):064114, 2019.

\bibitem{khaliullin2010graphite}
Rustam~Z Khaliullin, Hagai Eshet, Thomas~D K{\"u}hne, J{\"o}rg Behler, and
  Michele Parrinello.
\newblock Graphite-diamond phase coexistence study employing a neural-network
  mapping of the ab initio potential energy surface.
\newblock {\em Phys. Rev. B}, 81(10):100103, 2010.

\bibitem{podryabinkin2017active}
Evgeny~V Podryabinkin and Alexander~V Shapeev.
\newblock Active learning of linearly parametrized interatomic potentials.
\newblock {\em Comput. Mater. Sci.}, 140:171--180, 2017.

\bibitem{behler2007generalized}
J{\"o}rg Behler and Michele Parrinello.
\newblock Generalized neural-network representation of high-dimensional
  potential-energy surfaces.
\newblock {\em Phys. Rev. Lett.}, 98(14):146401, 2007.

\bibitem{behler2008metadynamics}
J{\"o}rg Behler, Roman Marto{\v{n}}{\'a}k, Davide Donadio, and Michele
  Parrinello.
\newblock Metadynamics simulations of the high-pressure phases of silicon
  employing a high-dimensional neural network potential.
\newblock {\em Phys. Rev. Lett.}, 100(18):185501, 2008.

\bibitem{deringer2018realistic}
Volker~L Deringer, Noam Bernstein, Albert~P Bart{\'o}k, Matthew~J Cliffe,
  Rachel~N Kerber, Lauren~E Marbella, Clare~P Grey, Stephen~R Elliott, and
  G{\'a}bor Cs{\'a}nyi.
\newblock Realistic atomistic structure of amorphous silicon from
  machine-learning-driven molecular dynamics.
\newblock {\em J. Phys. Chem. Lett.}, 9(11):2879--2885, 2018.

\bibitem{dragoni2018achieving}
Daniele Dragoni, Thomas~D Daff, G{\'a}bor Cs{\'a}nyi, and Nicola Marzari.
\newblock Achieving dft accuracy with a machine-learning interatomic potential:
  Thermomechanics and defects in bcc ferromagnetic iron.
\newblock {\em Phys. Rev. Mater.}, 2(1):013808, 2018.

\bibitem{zong2018developing}
Hongxiang Zong, Ghanshyam Pilania, Xiangdong Ding, Graeme~J Ackland, and Turab
  Lookman.
\newblock Developing an interatomic potential for martensitic phase
  transformations in zirconium by machine learning.
\newblock {\em Npj Comput. Mater.}, 4(1):1--8, 2018.

\bibitem{artrith2015grand}
Nongnuch Artrith and Alexie~M Kolpak.
\newblock Grand canonical molecular dynamics simulations of {Cu--Au} nanoalloys
  in thermal equilibrium using reactive ann potentials.
\newblock {\em Comput. Mater. Sci.}, 110:20--28, 2015.

\bibitem{chiriki2016modeling}
Siva Chiriki and Satya~S Bulusu.
\newblock Modeling of {DFT} quality neural network potential for sodium
  clusters: Application to melting of sodium clusters ({Na20} to {Na40}).
\newblock {\em Chem. Phys. Lett.}, 652:130--135, 2016.

\bibitem{chiriki2017neural}
Siva Chiriki, Shweta Jindal, and Satya~S Bulusu.
\newblock Neural network potentials for dynamics and thermodynamics of gold
  nanoparticles.
\newblock {\em J. Chem. Phys.}, 146(8):084314, 2017.

\bibitem{sosso2012neural}
Gabriele~C Sosso, Giacomo Miceli, Sebastiano Caravati, J{\"o}rg Behler, and
  Marco Bernasconi.
\newblock Neural network interatomic potential for the phase change material
  {GeTe}.
\newblock {\em Phys. Rev. B}, 85(17):174103, 2012.

\bibitem{artrith2011high}
Nongnuch Artrith, Tobias Morawietz, and J{\"o}rg Behler.
\newblock High-dimensional neural-network potentials for multicomponent
  systems: Applications to zinc oxide.
\newblock {\em Phys. Rev. B}, 83(15):153101, 2011.

\bibitem{artrith2016implementation}
Nongnuch Artrith and Alexander Urban.
\newblock An implementation of artificial neural-network potentials for
  atomistic materials simulations: Performance for {TiO2}.
\newblock {\em Comput. Mater. Sci.}, 114:135--150, 2016.

\bibitem{morawietz2016van}
Tobias Morawietz, Andreas Singraber, Christoph Dellago, and J{\"o}rg Behler.
\newblock How van der waals interactions determine the unique properties of
  water.
\newblock {\em P. Natl. A. Sci.}, 113(30):8368--8373, 2016.

\bibitem{cheng2016nuclear}
Bingqing Cheng, Jörg Behler, and Michele Ceriotti.
\newblock Nuclear quantum effects in water at the triple point: Using theory as
  a link between experiments.
\newblock {\em J. Phys. Chem. Lett.}, 7(12):2210--2215, 2016.

\bibitem{jose2012construction}
KV~Jovan Jose, Nongnuch Artrith, and J{\"o}rg Behler.
\newblock Construction of high-dimensional neural network potentials using
  environment-dependent atom pairs.
\newblock {\em J. Chem. Phys.}, 136(19):194111, 2012.

\bibitem{gastegger2016comparing}
Michael Gastegger, Clemens Kauffmann, J{\"o}rg Behler, and Philipp Marquetand.
\newblock Comparing the accuracy of high-dimensional neural network potentials
  and the systematic molecular fragmentation method: A benchmark study for
  all-trans alkanes.
\newblock {\em J. Chem. Phys.}, 144(19):194110, 2016.

\bibitem{boes2017neural}
Jacob~R Boes and John~R Kitchin.
\newblock Neural network predictions of oxygen interactions on a dynamic pd
  surface.
\newblock {\em Mol. Simulat.}, 43(5-6):346--354, 2017.

\bibitem{bartok2010gaussian}
Albert~P Bart{\'o}k, Mike~C Payne, Risi Kondor, and G{\'a}bor Cs{\'a}nyi.
\newblock Gaussian approximation potentials: The accuracy of quantum mechanics,
  without the electrons.
\newblock {\em Phys. Rev. Lett.}, 104(13):136403, 2010.

\bibitem{khorshidi2016amp}
Alireza Khorshidi and Andrew~A Peterson.
\newblock Amp: A modular approach to machine learning in atomistic simulations.
\newblock {\em Comput. Phys. Commun.}, 207:310--324, 2016.

\bibitem{wang2018deepmd}
Han Wang, Linfeng Zhang, Jiequn Han, and E~Weinan.
\newblock Deepmd-kit: A deep learning package for many-body potential energy
  representation and molecular dynamics.
\newblock {\em Comput. Phys. Commun.}, 228:178--184, 2018.

\bibitem{shapeev2016moment}
Alexander~V Shapeev.
\newblock Moment tensor potentials: A class of systematically improvable
  interatomic potentials.
\newblock {\em Multiscale Model. Sim.}, 14(3):1153--1173, 2016.

\bibitem{schutt2018schnetpack}
KT~Schutt, Pan Kessel, Michael Gastegger, KA~Nicoli, Alexandre Tkatchenko, and
  K-R Muller.
\newblock Schnetpack: A deep learning toolbox for atomistic systems.
\newblock {\em J. Chem. Theory Comput.}, 15(1):448--455, 2018.

\bibitem{desai2020implementing}
Saaketh Desai, Samuel~Temple Reeve, and James~F Belak.
\newblock Implementing a neural network interatomic model with performance
  portability for emerging exascale architectures.
\newblock {\em arXiv preprint arXiv:2002.00054}, 2020.

\bibitem{thompson2015spectral}
Aidan~P Thompson, Laura~P Swiler, Christian~R Trott, Stephen~M Foiles, and
  Garritt~J Tucker.
\newblock Spectral neighbor analysis method for automated generation of
  quantum-accurate interatomic potentials.
\newblock {\em J. Comput. Phys.}, 285:316--330, 2015.

\bibitem{bartok2013representing}
Albert~P Bart{\'o}k, Risi Kondor, and G{\'a}bor Cs{\'a}nyi.
\newblock On representing chemical environments.
\newblock {\em Phys. Rev. B}, 87(18):184115, 2013.

\bibitem{rupp2015machine}
Matthias Rupp.
\newblock Machine learning for quantum mechanics in a nutshell.
\newblock {\em Int. J. Quantum Chem.}, 115(16):1058--1073, 2015.

\bibitem{behler2011neural}
J{\"o}rg Behler.
\newblock Neural network potential-energy surfaces in chemistry: a tool for
  large-scale simulations.
\newblock {\em Phys. Chem. Chem. Phys.}, 13(40):17930--17955, 2011.

\bibitem{deringer2019machine}
Volker~L Deringer, Miguel~A Caro, and G{\'a}bor Cs{\'a}nyi.
\newblock Machine learning interatomic potentials as emerging tools for
  materials science.
\newblock {\em Adv. Mater.}, 31(46):1902765, 2019.

\bibitem{behler2017first}
J{\"o}rg Behler.
\newblock First principles neural network potentials for reactive simulations
  of large molecular and condensed systems.
\newblock {\em Angew. Chem. Int. Ed.}, 56(42):12828--12840, 2017.

\bibitem{ramprasad2017machine}
Rampi Ramprasad, Rohit Batra, Ghanshyam Pilania, Arun Mannodi-Kanakkithodi, and
  Chiho Kim.
\newblock Machine learning in materials informatics: recent applications and
  prospects.
\newblock {\em Npj Comput. Mater.}, 3(1):1--13, 2017.

\bibitem{behler2016perspective}
J{\"o}rg Behler.
\newblock Perspective: Machine learning potentials for atomistic simulations.
\newblock {\em J. Chem. Phys.}, 145(17):170901, 2016.

\bibitem{handley2010potential}
Chris~M Handley and Paul~LA Popelier.
\newblock Potential energy surfaces fitted by artificial neural networks.
\newblock {\em J. Phys. Chem. A}, 114(10):3371--3383, 2010.

\bibitem{szlachta2014accuracy}
Wojciech~J Szlachta, Albert~P Bart{\'o}k, and G{\'a}bor Cs{\'a}nyi.
\newblock Accuracy and transferability of gaussian approximation potential
  models for tungsten.
\newblock {\em Phys. Rev. B}, 90(10):104108, 2014.

\bibitem{bartok2015gaussian}
Albert~P. Bartók and Gábor Csányi.
\newblock Gaussian approximation potentials: A brief tutorial introduction.
\newblock {\em Int. J. Quantum Chem.}, 115(16):1051--1057, 2015.

\bibitem{huang2016communication}
Bing Huang and O~Anatole Von~Lilienfeld.
\newblock Communication: Understanding molecular representations in machine
  learning: The role of uniqueness and target similarity, 2016.

\bibitem{faber2018alchemical}
Felix~A Faber, Anders~S Christensen, Bing Huang, and O~Anatole Von~Lilienfeld.
\newblock Alchemical and structural distribution based representation for
  universal quantum machine learning.
\newblock {\em J. Chem. Phys.}, 148(24):241717, 2018.

\bibitem{christensen2020fchl}
Anders~S Christensen, Lars~A Bratholm, Felix~A Faber, and O~Anatole~von
  Lilienfeld.
\newblock {FCHL} revisited: Faster and more accurate quantum machine learning.
\newblock {\em J. Chem. Phys.}, 152(4):044107, 2020.

\bibitem{rupp2012fast}
Matthias Rupp, Alexandre Tkatchenko, Klaus-Robert M{\"u}ller, and O~Anatole
  Von~Lilienfeld.
\newblock Fast and accurate modeling of molecular atomization energies with
  machine learning.
\newblock {\em Phys. Rev. Lett.}, 108(5):058301, 2012.

\bibitem{chmiela2017machine}
Stefan Chmiela, Alexandre Tkatchenko, Huziel~E Sauceda, Igor Poltavsky,
  Kristof~T Sch{\"u}tt, and Klaus-Robert M{\"u}ller.
\newblock Machine learning of accurate energy-conserving molecular force
  fields.
\newblock {\em Sci. Adv.}, 3(5):e1603015, 2017.

\bibitem{zuo2019performance}
Yunxing Zuo, Chi Chen, Xiang-Guo Li, Zhi Deng, Yiming Chen, J{\"o}rg Behler,
  G{\'a}bor Cs{\'a}nyi, Alexander~V Shapeev, Aidan~P Thompson, Mitchell~A Wood,
  et~al.
\newblock A performance and cost assessment of machine learning interatomic
  potentials.
\newblock {\em J. Phys. Chem. A}, 2019.

\bibitem{deringer2018data}
Volker~L Deringer, Chris~J Pickard, and G{\'a}bor Cs{\'a}nyi.
\newblock Data-driven learning of total and local energies in elemental boron.
\newblock {\em Phys. Rev. Lett.}, 120(15):156001, 2018.

\bibitem{deringer2017machine}
Volker~L Deringer and G{\'a}bor Cs{\'a}nyi.
\newblock Machine learning based interatomic potential for amorphous carbon.
\newblock {\em Phys. Rev. B}, 95(9):094203, 2017.

\bibitem{hajinazar2017stratified}
Samad Hajinazar, Junping Shao, and Aleksey~N Kolmogorov.
\newblock Stratified construction of neural network based interatomic models
  for multicomponent materials.
\newblock {\em Phys. Rev. B}, 95(1):014114, 2017.

\bibitem{balabin2011support}
Roman~M Balabin and Ekaterina~I Lomakina.
\newblock Support vector machine regression ({LS-SVM})—an alternative to
  artificial neural networks ({ANNs}) for the analysis of quantum chemistry
  data?
\newblock {\em Phys. Chem. Chem. Phys.}, 13(24):11710--11718, 2011.

\bibitem{li2015molecular}
Zhenwei Li, James~R Kermode, and Alessandro De~Vita.
\newblock Molecular dynamics with on-the-fly machine learning of
  quantum-mechanical forces.
\newblock {\em Phys. Rev. Lett.}, 114(9):096405, 2015.

\bibitem{glielmo2017accurate}
Aldo Glielmo, Peter Sollich, and Alessandro De~Vita.
\newblock Accurate interatomic force fields via machine learning with covariant
  kernels.
\newblock {\em Phys. Rev. B}, 95(21):214302, 2017.

\bibitem{huan2019iterative}
Tran~Doan Huan, Rohit Batra, James Chapman, Chiho Kim, Anand Chandrasekaran,
  and Rampi Ramprasad.
\newblock Iterative-learning strategy for the development of
  application-specific atomistic force fields.
\newblock {\em J. Phys. Chem. C}, 123(34):20715--20722, 2019.

\bibitem{chapman2020machine}
J~Chapman, R~Batra, and R~Ramprasad.
\newblock Machine learning models for the prediction of energy, forces, and
  stresses for platinum.
\newblock {\em Comput. Mater. Sci.}, 174:109483, 2020.

\bibitem{batra2019general}
Rohit Batra, Huan~Doan Tran, Chiho Kim, James Chapman, Lihua Chen, Anand
  Chandrasekaran, and Rampi Ramprasad.
\newblock General atomic neighborhood fingerprint for machine learning-based
  methods.
\newblock {\em J. Phys. Chem. C}, 123(25):15859--15866, 2019.

\bibitem{isayev2017universal}
Olexandr Isayev, Corey Oses, Cormac Toher, Eric Gossett, Stefano Curtarolo, and
  Alexander Tropsha.
\newblock Universal fragment descriptors for predicting properties of inorganic
  crystals.
\newblock {\em Nat. Commun.}, 8(1):1--12, 2017.

\bibitem{xie2018crystal}
Tian Xie and Jeffrey~C Grossman.
\newblock Crystal graph convolutional neural networks for an accurate and
  interpretable prediction of material properties.
\newblock {\em Phys. Rev. Lett.}, 120(14):145301, 2018.

\bibitem{chen2019graph}
Chi Chen, Weike Ye, Yunxing Zuo, Chen Zheng, and Shyue~Ping Ong.
\newblock Graph networks as a universal machine learning framework for
  molecules and crystals.
\newblock {\em Chem. Mater.}, 31(9):3564--3572, 2019.

\bibitem{gilmer2017neural}
Justin Gilmer, Samuel~S Schoenholz, Patrick~F Riley, Oriol Vinyals, and
  George~E Dahl.
\newblock Neural message passing for quantum chemistry.
\newblock In {\em Proceedings of the 34th ICML-Vol. 70}, pages 1263--1272.
  JMLR. org, 2017.

\bibitem{jain2013commentary}
Anubhav Jain, Shyue~Ping Ong, Geoffroy Hautier, Wei Chen, William~Davidson
  Richards, Stephen Dacek, Shreyas Cholia, Dan Gunter, David Skinner, Gerbrand
  Ceder, et~al.
\newblock Commentary: The materials project: A materials genome approach to
  accelerating materials innovation.
\newblock {\em APL Mater.}, 1(1):011002, 2013.

\bibitem{ramakrishnan2014quantum}
Raghunathan Ramakrishnan, Pavlo~O Dral, Matthias Rupp, and O~Anatole
  Von~Lilienfeld.
\newblock Quantum chemistry structures and properties of 134 kilo molecules.
\newblock {\em Sci. Data}, 1:140022, 2014.

\bibitem{sun2019data}
Sheng Sun, Runhai Ouyang, Bochao Zhang, and Tong-Yi Zhang.
\newblock Data-driven discovery of formulas by symbolic regression.
\newblock {\em MRS Bulletin}, 44(7):559--564, 2019.

\bibitem{hernandez2019fast}
Alberto Hernandez, Adarsh Balasubramanian, Fenglin Yuan, Simon~AM Mason, and
  Tim Mueller.
\newblock Fast, accurate, and transferable many-body interatomic potentials by
  symbolic regression.
\newblock {\em Npj Comput. Mater.}, 5(1):1--11, 2019.

\bibitem{bianchini2016modelling}
F~Bianchini, JR~Kermode, and Alessandro De~Vita.
\newblock Modelling defects in {Ni--Al} with eam and dft calculations.
\newblock {\em Model. Simul. Mater. Sci.}, 24(4):045012, 2016.

\bibitem{pun2019physically}
GP~Purja Pun, R~Batra, R~Ramprasad, and Y~Mishin.
\newblock Physically informed artificial neural networks for atomistic modeling
  of materials.
\newblock {\em Nat. Commun.}, 10(1):1--10, 2019.

\bibitem{chen2019machine}
Lihua Chen, Huan Tran, Rohit Batra, Chiho Kim, and Rampi Ramprasad.
\newblock Machine learning models for the lattice thermal conductivity
  prediction of inorganic materials.
\newblock {\em Comput. Mater. Sci.}, 170:109155, 2019.

\bibitem{smith2018less}
Justin~S Smith, Ben Nebgen, Nicholas Lubbers, Olexandr Isayev, and Adrian~E
  Roitberg.
\newblock Less is more: Sampling chemical space with active learning.
\newblock {\em J. Chem. Phys.}, 148(24):241733, 2018.

\bibitem{zhang2019active}
Linfeng Zhang, De-Ye Lin, Han Wang, Roberto Car, and E~Weinan.
\newblock Active learning of uniformly accurate interatomic potentials for
  materials simulation.
\newblock {\em Phys. Rev. Mater.}, 3(2):023804, 2019.

\bibitem{musil2019fast}
Felix Musil, Michael~J Willatt, Mikhail~A Langovoy, and Michele Ceriotti.
\newblock Fast and accurate uncertainty estimation in chemical machine
  learning.
\newblock {\em J. Chem. Theory and Comput.}, 15(2):906--915, 2019.

\bibitem{jinnouchi2019phase}
Ryosuke Jinnouchi, Jonathan Lahnsteiner, Ferenc Karsai, Georg Kresse, and Menno
  Bokdam.
\newblock Phase transitions of hybrid perovskites simulated by machine-learning
  force fields trained on the fly with bayesian inference.
\newblock {\em Phys. Rev. Lett.}, 122(22):225701, 2019.

\bibitem{jinnouchi2019fly}
Ryosuke Jinnouchi, Ferenc Karsai, and Georg Kresse.
\newblock On-the-fly machine learning force field generation: Application to
  melting points.
\newblock {\em Phys. Rev. B}, 100(1):014105, 2019.

\bibitem{vandermause2019fly}
Jonathan Vandermause, Steven~B Torrisi, Simon Batzner, Alexie~M Kolpak, and
  Boris Kozinsky.
\newblock On-the-fly bayesian active learning of interpretable force-fields for
  atomistic rare events.
\newblock {\em arXiv preprint arXiv:1904.02042}, 2019.

\bibitem{loeffler2020active}
Troy~David Loeffler, Tarak~K Patra, Henry Chan, Mathew~J Cherukara, and
  Subramanian~KRS Sankaranarayanan.
\newblock Active learning the potential energy landscape for water clusters
  from sparse training data.
\newblock {\em J. Phys. Chem. C}, 2020.

\bibitem{stone1981distributed}
AJ~Stone.
\newblock Distributed multipole analysis, or how to describe a molecular charge
  distribution.
\newblock {\em Chem. Phys. Lett.}, 83(2):233--239, 1981.

\bibitem{mortier1986electronegativity}
Wilfried~J Mortier, Swapan~K Ghosh, and S~Shankar.
\newblock Electronegativity-equalization method for the calculation of atomic
  charges in molecules.
\newblock {\em J. Am. Chem. Soc.}, 108(15):4315--4320, 1986.

\bibitem{rappe1991charge}
Anthony~K Rappe and William~A Goddard~III.
\newblock Charge equilibration for molecular dynamics simulations.
\newblock {\em J. Phys. Chem.}, 95(8):3358--3363, 1991.

\bibitem{dammak2009quantum}
Hichem Dammak, Yann Chalopin, Marine Laroche, Marc Hayoun, and Jean-Jacques
  Greffet.
\newblock Quantum thermal bath for molecular dynamics simulation.
\newblock {\em Phys. Rev. Lett.}, 103(19):190601, 2009.

\bibitem{binder2011monte}
Kurt Binder, Benjamin Block, Subir~K Das, Peter Virnau, and David Winter.
\newblock Monte carlo methods for estimating interfacial free energies and line
  tensions.
\newblock {\em J. Stat. Phys.}, 144(3):690--729, 2011.

\bibitem{reeve2017error}
Samuel~Temple Reeve and Alejandro Strachan.
\newblock Error correction in multi-fidelity molecular dynamics simulations
  using functional uncertainty quantification.
\newblock {\em J. Comput. Phys.}, 334:207--220, 2017.

\bibitem{marrink2007martini}
Siewert~J Marrink, H~Jelger Risselada, Serge Yefimov, D~Peter Tieleman, and
  Alex~H De~Vries.
\newblock The {MARTINI} force field: coarse grained model for biomolecular
  simulations.
\newblock {\em J. Phys. Chem. B}, 111(27):7812--7824, 2007.

\bibitem{onufriev2018water}
Alexey~V Onufriev and Saeed Izadi.
\newblock Water models for biomolecular simulations.
\newblock {\em Wiley Interdiscip. Rev. Comput. Mol. Sci.}, 8(2):e1347, 2018.

\end{thebibliography}

\end{document}